\documentclass[aps,showpacs,preprintnumbers,amsmath,amssymb,twocolumn, tightenlines]{revtex4}

\usepackage{graphicx}
\usepackage{bm}
\usepackage{amsmath}
\usepackage{dcolumn}
\usepackage{bm}

\newcommand{\be}{\begin{eqnarray}}
\newcommand{\ee}{\end{eqnarray}}

\begin{document}
\title{The Rise and Fall of Baryons }

\author{ Shu Lin and Edward Shuryak}

\affiliation{ 
Department of Physics and Astronomy, Stony Brook University, 
Stony Brook NY 11794-3800, USA
}

\date{\today}

\vspace{0.1in}
\begin{abstract}
We discuss the baryonic contribution to QCD thermodynamics near the QCD phase transition, which we 
split into ``stringy excitations'' and the ``chiral'' lowest states. Our finite-$T$ 
 string model for the former component is  inspired by the lattice data on static potentials and ideas of string survival even $above$ $T_c$:
it is  used to explain two sets of  baryonic susceptibilities calculated on the lattice by the Bielefeld-BNL group.
Two new ingredients of the model are (i) the near-$T_c$ tightening of the strings, and (ii) physical upper limit of the string length. 
Then we proceed to more subtle effects  related to chiral restoration dynamics near $T_c$:  we suggest that ``melting'' of the sigma terms 
of the lowest nucleon/$\Delta$ masses 
can explain these susceptibilities. In a discussion, we consider bound monopole-quark states, as a possible explanation to
some deficits of contribution at $T=(1-1.4)T_c$.
\end{abstract}
\maketitle

\section{Introduction}
  The issues discussed in this paper deal specifically with the sector of the QCD states with a nonzero baryonic charge, the 
  $baryons$ in the confining phase and $quarks$ in the deconfined one. They are however 
part of a  more general discussion about  the fate of hadronic states
  near and above the deconfinement phase transition.
 
Hadronic spectroscopy  studies the lowest hadronic states  in great details:
but as the excitation energy grows the task of exact quantum number identification gets more challenging. Eventually the
decay widths of the hadrons become comparable to distances between the states, or exceeds it $\Gamma > |E_i-E_j|$: here identification
of the individual states becomes impossible. Yet it does not mean that the states (and their density) do not exist: multiple examples from other fields of physics -- e.g.
the overlapping neutron resonances in nuclear physics -- show that.  The thermodynamical quantities are averaged over all states excited at 
some well defined conditions, and those include the states which we cannot identify individually. We need certain models about
hadronic states at one hand, to understand/reproduce the available (lattice) results on the thermodynamics: this is the main logical
direction of this paper.

 While  the temperature and the baryonic chemical potential are small  $T,\mu\ll T_c$, hadronic matter is a dilute gas of basic hadrons, e.g. pions and nucleons, and thermodynamics must be  
 given by elementary ideal gas formulae, plus interaction corrections given by the standard virial expansion.
  As one approaches $T_c$ from below, many excited states of basic hadrons  come into play, and  hadrons
become more densely packed and more strongly interacting. It is widely believed that a consistent treatment
of this matter is given by a ``resonance gas'' approximation,  in which interaction is taken care of by counting all resonances on equal footing
with stable hadrons\cite{tawfik}. Furthermore, when the vicinity of the critical region is reached
$T\sim T_c= 170-190\, MeV$, new phenomenon (referred to as {\em ``rise''} in the title) happens, in which
thermodynamical quantities grow by about an order of magnitude in a rather narrow interval of temperatures.

Statistical bootstrap model (hadrons are bags made of other bags) lead Hagedorn in 1960's to the conclusion that the density of hadronic states
$\rho(m)$ may grow exponentially with the mass $m$. If this be the case, Hagedorn argued,  at some temperature this growth would be cancelling the Boltzmann exponent,  preventing hadronic systems from being heated to
temperatures above it. Thus he predicted  certain {\em maximal possible temperature} in Nature, $T<T_{Hagedorn}$
\cite{hagedorn}. The development of QCD in 1973 has lead to the opposite conclusion:  the high-$T$ phase of QCD is in fact  the so called {\em quark-gluon plasma} (QGP)
\cite{QGP} in which the density of states grows like the power of the mass, like for other ``normal'' systems.
 Later,  Hagedorn reinterpreted  his argument  as the maximal possible temperature of the {\em 
hadronic phase}, taking it as an early indication to the existence of the QCD deconfinement  phase transition. 
The basic physics of Hagedorn phenomena
was   explained by Polyakov and Susskind in 1970's who pointed out that exponential growth
of the density of states is the natural consequence of the QCD string model.

As $T$ gets  higher than $T_c$, hadronic (e.g. baryonic) contribution to thermodynamics  
 gets replaced by free quarks and gluons.. The issues we will be discussing
  in this paper is what exactly happens in the so called ``RHIC domain'', $T=(1-2)T_c$. As we will see,
  lattice susceptibilities tell us that the
rapid rise of hadronic states changes to nearly equally rapid {\em fall}, at $T\sim 1.1 T_c$:
there is no consensus about its exact physical nature. This question will in turn lead us to the issue of whether
 the very existence of strings above $T_c$ is or is not possible. Our point of view is that not only 
 it is, but recent RHIC discovery of the so called ``soft ridge'' shows existence of quite robust strings,
 surviving for several fm/c under such conditions, for related discussion see  \cite{Shuryak:2009cy}.  We also would claim that ``stringy''
 excitations dominate the rise and the fall of the baryonic excitations: the latter related to the increase of quark density and
 shortening of the $mean$ length of the strings. 
 
In order to discuss these issues and provide some (model dependent) answers to those and many similar questions 
we had selected a particular subset of hadrons, namely the $baryons$.  The  reason for that  is that
by considering susceptibilities -- the quantities obtained by certain number of derivatives over the quark chemical potential $\mu$ -- we insulate the issue from 
much more complicated
 sector of states with
$zero$ baryon number, including mesons, glueballs or even more exotic quasiparticles such as monopoles/dyons. 

The  inputs
to our study is provided by  the lattice data sets from
two papers, by UK-Bielefeld-BNL group,to be referred below as the Set 1 and Set 2, respectively. The former 
one \cite{allton} is for $N_f=2$, two  quark flavors  with a rather heavy
pion mass, $m_\pi\approx 770\, MeV$. The 
 latter is a more recent work \cite{cheng} is closer to the real world,  it includes the strange quark and also has much smaller
pion mass  $m_\pi\approx 220\, MeV$. (Unfortunately, the latter one has also somewhat larger fluctuations and thus error bars.)
Since those two sets
correspond to two different worlds,  all quantities in question -- including the baryon masses and the critical temperatures themselves -- are different,
and reproduction of their  differences would be one of the tasks  for our model.

\begin{figure}[t]
\vspace{0.75cm}
\includegraphics[width=8cm]{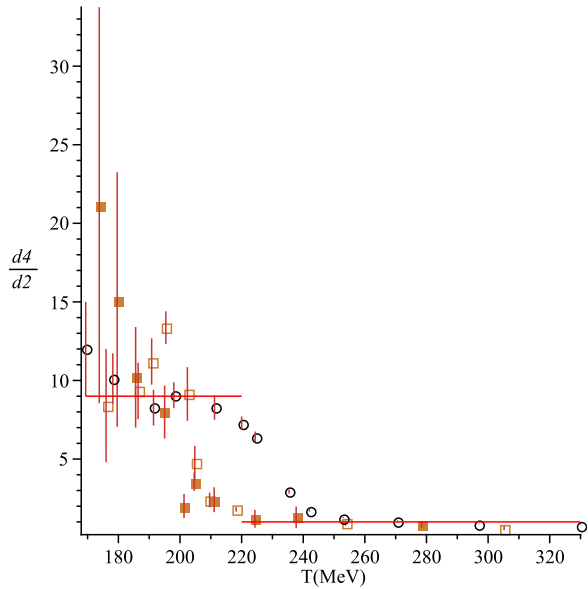}
\caption[h]{\label{fig_kur}(Color online)The temperature dependence of the ``kurtosis'',  the ratio of susceptibilities $d_4/d_2$. The Set 1
(2 flavor QCD at $m_{\pi}=770MeV$)  are shown by black circles, while Set 2 (2+1 flavor QCD at 
$m_{\pi}=220MeV$) is shown by (brown) solid boxes for $N_\tau=6$ and open boxes for
$N_\tau=4$). The upper and lower horizontal lines  correspond to the values  $9$ and $1$, respectively.}
\end{figure}

 The most prominent feature of these susceptibilities are illustrated in Fig.\ref{fig_kur}, 
  which shows the ratio of the 4-th to the second derivative over quark chemical potential, to be
called  $kurtosis$ for brevity.  One can see from this figure  that the
kurtosis drops rapidly from $9$ to $1$ as the temperature crosses $T_c$. 
It finds a very simple explanation: while the baryonic partition function is proportional to $\cosh(3\mu/T)$, that of quarks has just  $\cosh(\mu/T)$,
and two derivatives of those functions explain $9$ and  $1$ in the corresponding phases. This means that the confined phase is dominated by baryons and
the deconfined one by quarks. 

 While this conclusion is basically correct, a detailed study of these transition and the $T$-dependence of the susceptibilities  will reveal many more interesting
 details of the transition. The susceptibility $d_4$ in particular show a very dramatic rise at $T_c$, complemented by a rapid drop right after that. 
 Both these features are driven by the baryon sector, and not by the quarks -- as we emphasized by the title of the paper. 
  The reason for this to happen, as well as some model describing it, is the main part of this paper.

 Physics related to these data has been discussed by Liao and Shuryak \cite{Liao}, who emphasized the presence of at least the lowest baryons 
 in the region of $T=(1-1.4)T_c$, and in particular the $T,\mu$-dependence of their masses.
 It was based on a previous paper by the same authors  \cite{Liao:2005hj} in which
  lattice-based potentials between quarks were used to study when quark quasiparticles can form bound states.
  From a variational solution to the
  3-quark Schr\"{o}dinger equation they concluded that only the lowest states should survive, remaining  bound in the $T$ interval just mentioned.
  Another issue is whether the baryon and quark effective masses are $\mu$-dependent. If they are, that generates deviations from     $9$ and $1$
  in the kurtosis, and certain changes in the susceptibilities: as we will show below, those compare well with the latest lattice data.
  In our discussion below we will separate those lowest states from ``stringy'' excitations. 
   
  Weise and collaborators \cite{Ratti:2005jh,Roessner:2006xn,Ratti:2007jf} have  provided
very nice descriptions of these data in the PNJL model: their main idea was that the single  quark and diquark contributions are both suppressed
close to $T_c$
by a small VEV of the Polyakov loop $\langle L(T)\rangle$. 
Although this model has no baryons per se, only the colorless combinations of three quarks (so to say,  ``proto-baryons'') are 
the colorless states which are not suppressed by the Polyakov loop in the near-$T_c$ region, and are thus the main contributor to thermodynamics there.


   Finally, there are two more important physics issues, both related to the chiral dynamics. The first is the dependence
  of baryonic susceptibilities (and of course masses) on the value of the light quark mass $m_q$ in the theory. 
  It has been studied a lot by the lattice community, but in the vacuum (at $T=0$) 
  rather than at  $T\approx T_c$. The second is the derivative
  of the baryon mass over the quark mass -- known as the ``sigma term''. It is a chirally odd quantity, expected to
  vanish at $T>T_c$ in the chiral restoration transition. So, we will be look for observable consequences of that, in the lattice data set under consideration.
  

The paper is structured as follows. Section \ref{sec_stringy} is devoted to ``stringy excitations'' of baryons, in it we develop a somewhat modernize model
of the phenomenon, which goes back to Polyakov-Susskind excitation of strings. We will show that this phenomenon can indeed naturally describe very rapid rise
in baryonic contribution to thermodynamics near $T_c$. We have also 
 added a restriction on the upper limit of the string length
related to quark density: this turns out to produce the observed ``fall'' of the stringy baryons. The density of quarks in the near-$T_c$ region
is calculated with recent lattice data on quark effective masses as well as Polyakov loop, see   section \ref{sec_quarks}.
In section \ref{sec_results} we subtract the quark contribution from total susceptibilities and compare it to our model calculation
for ``stringy'' and ``non-stringy'' baryons.   Effects of chiral dynamics, which determine $m_q$, $T$ and $\mu$ dependencies of the baryon mass,
are discussed  in section \ref{sec_chiral}.  Discussion of some open issues is placed in section \ref{sec_discussion}: one of them is
the role of bound states between quarks and magnetic objects, monopoles/dyons, which (to our knowledge) are discussed in the
QGP context for the first time. We present some crude estimates of their possible contribution, and find that it seems to be
of the same magnitude  as the ``unaccountable'' contribution with quarks and baryons. 

\section{Stringy baryons} \label{sec_stringy}

\subsection{Hagedorn transition and the string model}

 Specific realization of Hagedorn's ideas came with
the advent of the QCD strings.   Let us briefly remind this well known argument here,   for introduction of notations. 
An important parameter of the string $a$, to be referred to as {\em ``the string scale''},
is loosely speaking a length of the string section which can be moved independent from other segments. (This loose idea implies that
it should be some $O(1)$ number times the string width.)  Its precise definition is given in terms of the
 number of string states.
Suppose a string has certain
length $L$ and tension $\sigma$, so its energy is
$E=\sigma L$. 
\be \rho(L) \sim exp(S(L))\sim 5^{L/a} \ee  
Here 5=2d-1 is the number of directions the string may turn to in d-dimensional space without self-crossing, after a segment of  length equal to $a$.
Now, combining the entropy and energy together one gets stringy partition function in the generic form
\be Z_{string} =\int dL e^{ln(5)L/a-\sigma L/T} \ee
which diverges as the temperature approaches the Hagedorn critical temperature
\be  T_c=\sigma a/ln(5) \ee
Note that at this temperature the exponent (the free energy of the string $F=E-TS$) goes to zero, allowing the string to get arbitrarily long. 

Now, does this simple string model describe properties of the deconfinement transition observed on the lattice, for various gauge theories?

Before answering this question, one needs to specify the string scale scaling. The string tension scales as \be \sigma\sim E^2 a^2\ee
if $E$ is the electric field strength and $a$ is the string width. At the same time the electric flux, which should match the electric
charge at the end of the string, is $ flux=Ea^2=const $, from which one concludes that the scaling
must be
\be \sigma \sim E\sim 1/a^2 \ee
Using this argument one expects that $T_c/\sqrt{\sigma}$ should be some universal constants: and in fact for pure gauge theories with different
number of colors $N_c$ this prediction actually works well for large $N_c$, see e.g. \cite{Teper}.

\begin{figure}[!t]
\vspace{0.75cm}
\includegraphics[width=8cm]{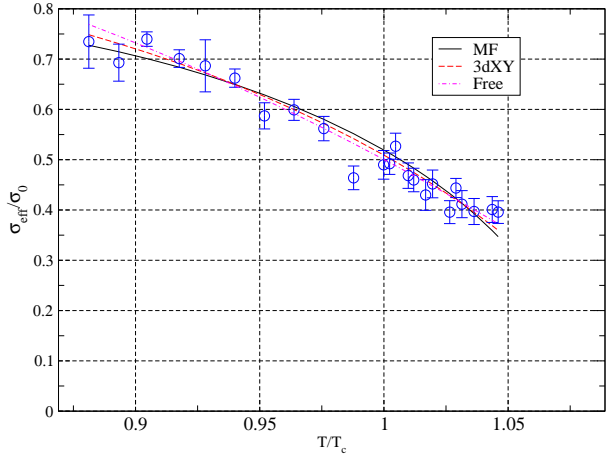}
\caption[h]{\label{fig_tensionsu10} (Color online) Temperature dependence of the free energy tension $\sigma_F$, for the SU(10) gauge theory according to Ref.\protect\cite{Teper}}
\end{figure}

For readers who are not simply satisfied by the scaling of $T_c$ and are unsure if the deconfinement is the string Hagedorn
transition, it is instructive to look at lattice measurements of the string free energy. Such measurements, for $N_c$ as large as 12 has been made
by Bringoltz and Teper \cite{Teper}: see one of their figures is reproduced in Fig.\ref{fig_tensionsu10}. As one can see from it,
with increasing $T$ the effective (free energy) tension is indeed decreasing, roughly linearly: this would agree with the idea
that the energy of the string is hardly affected by the temperature and the decrease is due to $T$ times the entropy of string states. However
 at $T/T_c=1$ the free energy tension
is $not$ zero but only about half of the vacuum value: using metastability of the 1-st order transition the authors were able to extend the curve 
about 5\% above $T_c$ and show that the trend continues: the projected Hagedorn temperature is perhaps $T_H=1.11T_c$ or so. Thus,
the deconfinement at large $N_c$ is $not$ the Hagedorn transition, although it scales similarly and  is numerically close to it. 

\subsection{Strings at finite temperatures in QCD-like theories with quarks}
 QCD with the physical quark masses is different from large $N_c$ pure gauge theories because instead of the first order transition with a large jump
 it has instead rather
 continuous ``crossover'' transition. The continuity of the transition allows us to ask if any role is played in the QCD phase transition  by
the Polyakov-Susskind string argument outlined above, at and above $T_c$. 
 
 Before we start answering this question, let us first review what is known about QCD strings, at zero and  finite-$T$, $microscopically$.
 At $T=0$ there are extensive lattice studies of string tension, shape, internal structure \cite{Bali98} and even the string breaking \cite{Bali05}. It is done by inserting two static charges and studying their influence on the gauge fields: it helps by the fact that at $T=0$ the string is just straight, with small influence of quantum vibrations.
   We will not go into details here, just mention that the shape fits well into the ``dual superconductor'' picture, with condensed monopoles creating a coil of magnetic current
   around it. The distribution of the electric field fits well into the solution of (dual) Maxwell equations in superconductor found by  Abrikosov
   \be E(r)\sim K_0(r/\lambda)\ee 
 with the length parameter $\lambda\approx .15 \, fm \approx (1.3 GeV)^{-1} $ \cite{Bali98}. 
 Unfortunately, at finite $T$ thermal fluctuations of the string make such studies much more challenging,
 and so we do not have similar results in the near-$T_c$ region.
 
 Yet there are still extensive lattice studies of the thermodynamical quantities associated with a pair of  two stationary $\bar q q$ 
 \cite{KZ, KZ2}. We will not review this large subject here, but only comment on the
  $linear$ (in distance) part of these static potentials. 
Let us define  the free energy
tension $\sigma_F(T)$ and entropy tension $\sigma_S(T)$ as follows

\be
&&V=\sigma_V(T)L+V_0 \\
&&F=\sigma_F(T)L+V_0 \\
&&S=\sigma_S(T)L
\ee
where $V_0$ is a constant. 
Standard thermodynamical relations $U=F+TS$, $S=-dF/dT$ 
imply the same relations for tensions
\be
\label{eqn_sigmav}
&&\sigma_V=\sigma_F+T\sigma_S \\
\label{eqn_sigmas}
&&\sigma_S=-d\sigma_F/dT
\ee
The free energy string tension $\sigma_F$, 
obtained by solving (\ref{eqn_sigmav}) and (\ref{eqn_sigmas}) with the string tension of the 
potential energy as the input:

\be\label{sigmaF}
\sigma_F=-T\int_{T_0}^T\frac{\sigma_V(T)}{T^2}
\ee

\begin{figure}[!ht]
\vspace{0.75cm}
\includegraphics[width=8cm]{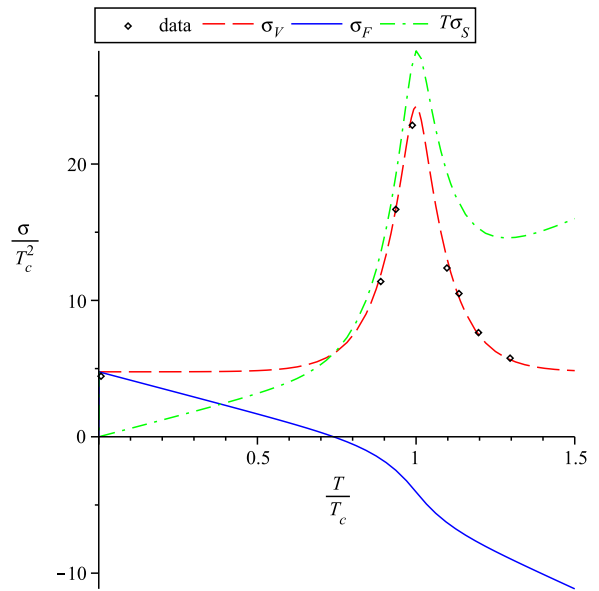}
\caption[h]{\label{fig_string_tension}(Color online) The temperature dependence of two string tensions, 
$\sigma_V$(red dashed line), $\sigma_F$(blue solid line). The entropy term
$T\sigma_S$ is shown by the (green dash-dotted line). The lattice data points shown are for $\sigma_V$ extracted from lattice work \cite{KZ2}. 
The constant used in the plot is $T_0=0.95T_c$. Its determination will be explained later.
}
\end{figure}

At $T_0$, the free energy tension vanishes, so it is the Hagedorn temperature
 of QCD, which will be fixed later.
In Fig.(\ref{fig_string_tension}) we show the input string tension of the
 potential energy and the resultant free energy string tension.  The input are
 lattice  $\sigma_V$ extracted from lattice work \cite{KZ2} parametrized by:

\be
{\sigma_V \over T_c^2}=26.21e^{\left (-11.40\sqrt{(T/T_c-1)^2+6.78\times 10^{-4}}\right )}+4.76.
\ee

\subsection{T-dependent string model and density of states}
One can build a simple model based on the same number of state as for the fundamental string. Suppose $T$-dependent string
scale $a(T)$ be known: then we postulate the density of states to be

\be  
\rho(L)\sim 5^{L/a(T)}=e^{\sigma_S(T) L}\label{length_density}
\ee  

but now with effective $T$-dependent string scale. To write the density 
of states per unit mass, we note $m=m_0+\sigma_V(T)L(T)$, with $m_0$ the 
contribution from the quarks at he ends of the string. 
As the string tension $\sigma_V(T)$ 
changes with temperature, the string length $L(T)$ corresponds to certain
baryon state changes qualitatively in the opposite way, i.e. a tight
string is short in length. As a result, the quark contribution will also
change with temperature. In order to take into account this effect, we
introduce a factor $x$ in the exponent. We further assume it is a constant
in the temperature interval we explore:

\be\label{mass_density}
&&\rho(m)|_{T=0}=ce^{\sigma_S L}=ce^{\sigma_S/\sigma_V(m-m_0)}\rightarrow\nonumber\\ 
&&\rho(m)|_{T>0}=ce^{x\sigma_S(T)/\sigma_V(T)(m-m_0)}
\ee

The normalization constant $c$ in (\ref{mass_density}) will be fixed by matching the experimental data at $T=0$.

In order to make comparisons with the output of lattice simulation, where large 
unphysical quark masses are used. We adopt the following parametrization for the baryon masses:

\be\label{b_mass}
m_B=\sigma_V(T)L(T) +N_s m_s+ km_{\pi}^2
\ee
in which the first string term is complemented by two more terms, which
describe the dependence of the ``string's ends'' on the strange and light quark masses.

The second term is just additive dependence on
the strange quark mass, with $N_s$ the number of strange quarks in the baryon.  
  The light 
quark mass (pion mass) dependence is included via the $slope$ \be k= {\partial m_B\over \partial m^2_\pi }\ee
known also as a ``sigma term''.  It depends nontrivially on the baryon in question, see
lattice studies such as
 \cite{lee}. The $N$ and $\Delta$ channels have
stronger dependence than the $\Lambda$ and $\Sigma$ channels, which contain 
more strange quarks. 
We list the slope $k$ extracted from 
\cite{lee} for relevant channel in Tab.\ref{tab:slope}. The $\Omega$ channel contains only strange quarks, therefore
we assume the slope to be zero. Note the lowest states of each channel is separated from
the other resonances by a significant gap, and they have different properties
(we will call them ``chiral baryons'' and further justify this separation 
in section \ref{sec_chiral}). 
 These lowest states (for three flavors
the 56-plet, spin 1/2 octet $N(938)$,$\Lambda(1116)$,
$\Sigma(1195)$,$\Xi(1317)$ and 3/2 decuplet $\Delta(1232)$,$\Sigma^*(1385)$,$\Xi^*(1530)$,$\Omega(1672)$, and for two flavors, the 20-plet of $N$ and $\Delta$) are treated separately from the rest of ``stringy'' (also called hybrid) excitations.

\begin{table}
\caption{\label{tab:slope}slope $k$ for different channels}
\begin{tabular}{ccccc}
$N(GeV^{-1})$& $\Delta(GeV^{-1})$& $\Lambda(GeV^{-1})$& $\Sigma(GeV^{-1})$& $\Xi(GeV^{-1})$\\
\hline
0.79& 0.73& 0.61& 0.55& 0.3
\end{tabular}
\end{table}

\begin{figure}[!t]
\vspace{0.75cm}
\includegraphics[width=7cm]{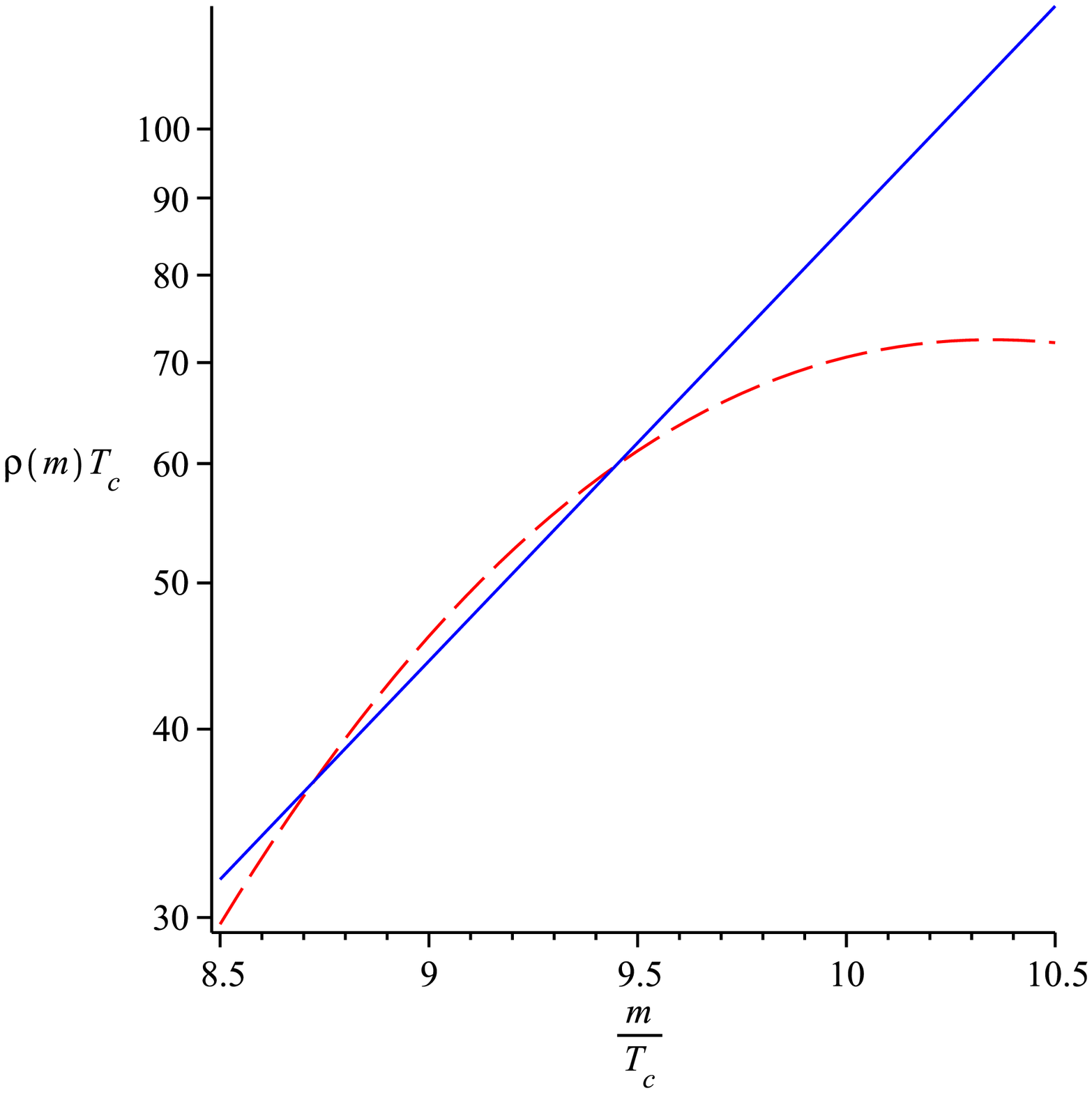}\\
\includegraphics[width=7cm]{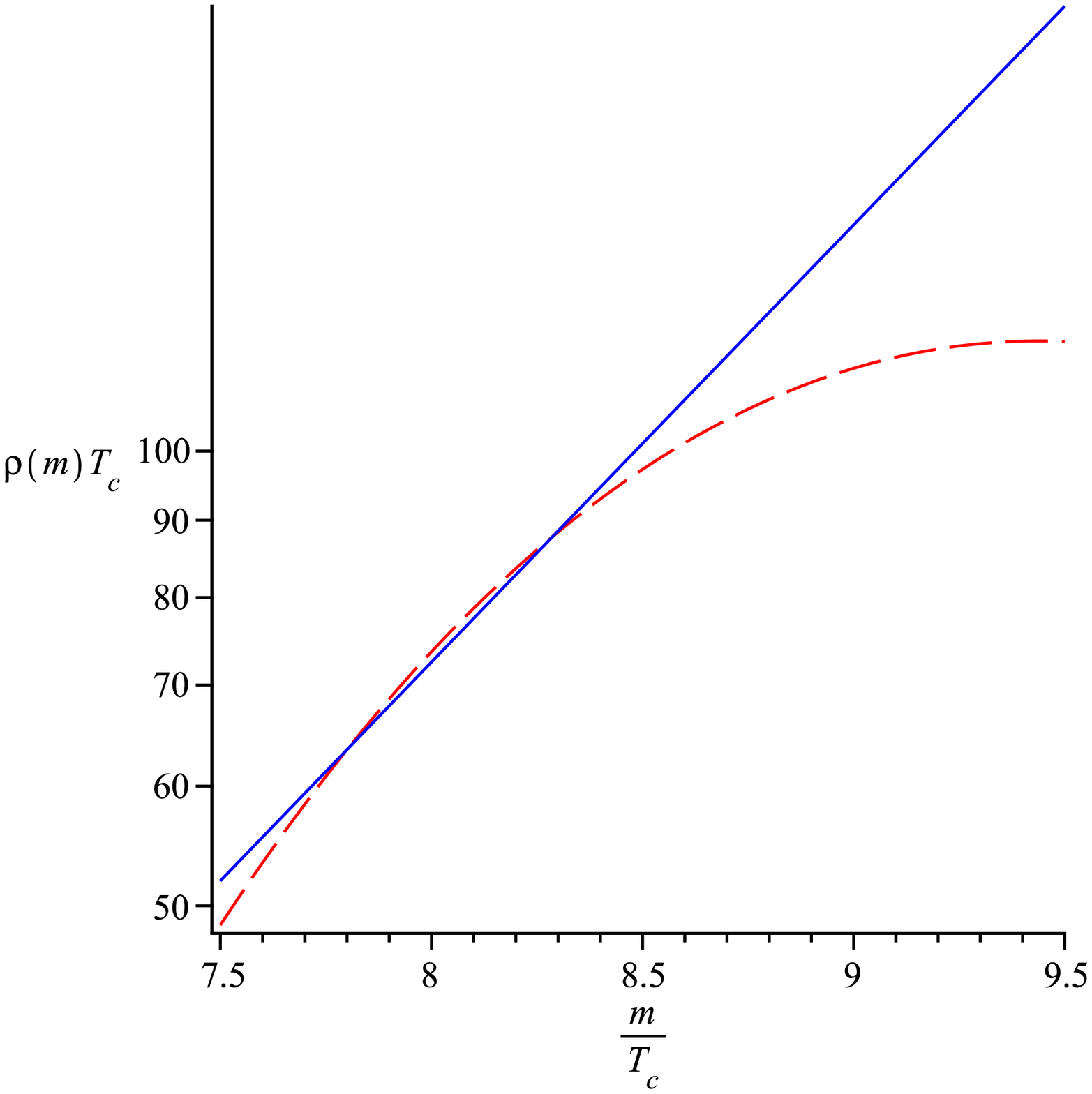}
\caption[h]{\label{fig_dos}(Color online) The ``string inspired'' exponential fit to the baryonic density of states, for Set 1 and Set 2
(upper and lower figure, respectively). The (red) dashed line is the sum of 
 over all 
  resonances listed in Particle Data Book \cite{PDG}, with masses corrected to  lattice quark mass by expression(\ref{b_mass}). 
The (blue) solid line is the exponential fit: $\rho(m)=0.11e^{m/1.5}$(Set 1)
and $\rho(m)=0.35e^{m/1.5}$(Set 2), with
unit set properly by $T_c$.}
\end{figure}

Let us return to the density of states of ``stringy excitations'', by which we count all resonances
to which, for smoothness we assign a width of
$200MeV$. 
We modify the baryonic resonances according to (\ref{b_mass}) and fit the 
density of states by the string inspired exponential function 
(\ref{mass_density}). 
Fig.\ref{fig_dos} shows the exponential fitting to the densities of states.

Comparing our stringy density of states with lattice modified baryonic resonances, we have the following parametrizations
for $N_f=2$ $c=0.11e^{m_0/1.5}$(Set 1) and $N_f=2+1$
$c=0.35e^{m_0/1.5}$(Set 2). From (\ref{b_mass}), we should have 
$m_0=km_{\pi}^2+N_sm_s$, which varies in different channels. Since we have counted
all stringy baryons indiscriminately in the density of states for statistical reasons. We simply use
the smallest $m_0$ among all channels.
Furthermore the Hagedorn temperature $T_0$ in will be
fixed from the relation $\sigma_S/\sigma_V=1/1.5T_c^{-1}$ at $T=0$, 
giving $T_0=0.95T_c$.
 It is important to note the critical temperature
corresponding to Set 1(2 flavors at $m_{\pi}=770MeV$) and Set 2(2+1 flavors at
$m_{\pi}=220MeV$) are slightly different. 
The former has $T_c=220MeV$, while the latter has $T_c=200MeV$(They can be 
obtained from\cite{mq_Tc,miao}). 
Special attention should be paid in converting the absolute unit to units of $T_c$,
different for each set.

\section{Quarks  in the near-$T_c$ region}
\label{sec_quarks}

We postulate that the free energy (pressure) due to quarks and antiquarks is given by the usual free gas formula
\be\label{p_quark}
p_q(T,\mu)=d\frac{1}{\pi^2}m_q^2T K_2(m_q/T)\cosh(\mu/T)\langle L(T)\rangle
\ee
modified by the last factor, 
 the so called the Polyakov loop,
\be 
\langle L\rangle={1\over N_c}Tr\left[{\cal P}e^{ig\int d\tau A_0(x)}\right]
\ee
induced by the presence of nonzero $A_0$ gauge fields.
The corresponding lattice data and our parametrization of both the assumed quark effective mass and the Polyakov factor are described in the Appendix.

As those calculations are rather standard, we would not show the quark contribution itself, showing in several plots below only the susceptibilities
with and $without$ quark contribution, shown by boxes and circles, respectively. 
The latter contributions show a near-$T_c$ peak and then decrease, disappearing around $1.5T_c$. Our task in the section to follow
would be the interpretation of
these ``non-free-quark'' contributions. 


As this and similar examples show, one has inevitably $\sigma_F(T)$ crossing zero. We would however argue that unlike in the string-Hagedorn 
transition, the string length should not be allowed to become indefinitely large and the partition function to diverge. The reason for that
is  string breaking/flipping. Let us introduce the {\em quark scale} 
\be\label{cubic} L_q(T) =n_q^{-1/3} \ee
related to their density in the QCD matter at temperature $T$. 

Note the quark density $n_q$ from 
naive derivative of the quark pressure with respect to $\mu$ will give 
$\sinh(\mu/T)$, which vanishes at $\mu=0$ due to the cancellation between
quarks and antiquarks. We instead use $\cosh(\mu/T)$, as both quark and
antiquark can be the end points of a string. It is also required by the
$\mu\leftrightarrow-\mu$ symmetry.
The degeneracy factor $d=2(2S+1)N_cN_f$. 
Suppose we start with a string, whose ends are fixed, for simplicity at some
nearby points. If the string length is $L$ and its shape is random, the average distance from the original ends would
be given by random walk expression
\be  <(\delta x)^2>\sim a^2 (L/a) \ee
where $L/a$ is the number of random turns. We will then argue that as a string reaches the ``domain of influence'' of other quarks,
the string would not grow anymore in length but get switched to another quark. We thus postulate that
\be <(\delta x)^2>\sim a^2 (L/a)  < L_q^2\ee
sets the upper limit in the integral over the string length: 
\be
L_{max}= y^2L_q^2/a
\ee
with $y$ is some constant of order 1. Correspondingly, $m_{max}=m_0+\sigma_V(T)L_{max}$.
The temperature dependence of $L_q$ is such that at $T<T_c$ it is very large, reaching large but finite value at $T_c$, and then rapidly decreasing.

\begin{figure}[!ht]
\vspace{0.75cm}
\includegraphics[width=6cm]{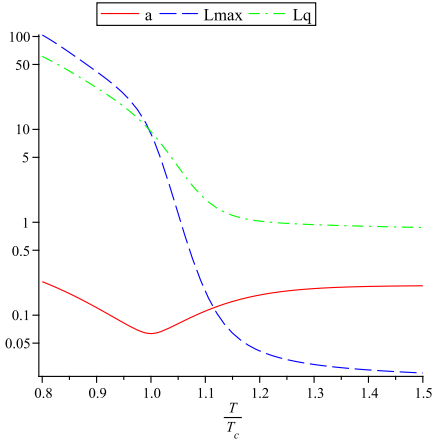}
\caption[h]{\label{fig_lma}(Color online)A comparison of the string length 
cutoff $L_{max}$(blue dashed line) and
string scale $a$(red solid line) at different temperatures for two flavors QCD. 
$L_q$(green dash-dotted line) is also included
for comparison. The parameters used in the plot is $y=0.08$, $T_0=0.95T_c$}
\end{figure}

We compare various length scale appearing in the model in Fig.\ref{fig_lma}. 
At temperature $T\sim 0.8T_c$, the string length can get very long, but long
strings are suppressed by Boltzmann factor. As the temperature crosses the
critical value, the density of states starts to win over the Boltzmann factor
and the string length cutoff becomes important, as illustrated by the dropping
of the $L_{max}$ near $T_c$. At $T\sim 1.1T_c$, the stringy baryon picture
breaks down as $L_{max}\sim a$.

\section{Thermodynamics of the ``stringy baryons'' in the near-$T_c$ region}\label{sec_results}

Now we collect all the pieces of our string-inspired model and calculate several susceptibilities $d_n(T)$, to be compared with
Set 1 and 2 lattice data. The expressions of the pressure and susceptibilities
due to baryons are shown in the following:



\be
\label{string_p}
&&p(T,\mu)\equiv\frac{T\partial \ln Z(T,\mu)}{\partial V}=\sum_{i}\frac{1}{\pi^2}\frac{m_i^2}{T^2}K_2(m_i/T)cosh(3\mu/T)\nonumber\\
&&+\int_{m_{min}}^{m_{max}}\frac{1}{\pi^2}\frac{m^2}{T^2}K_2(m/T)cosh(3\mu/T)\rho(m)dm\\
\label{string_dn}
&&d_n(T)=\frac{\partial^n(p/T^4)}{\partial(\mu/T)^n}\vert_{\mu=0}
\ee
The lowest states counted separately by the index $i$ include the 20-plet of $N,\Delta$ for $N_f=2$ 
or the members of the 56-plet,
if the strange quark is present.
The second term is the contribution from the
  continuum assumed to be stringy excitations. The upper bound in integration is given by
$m_{max}=m_0+\sigma_V(T)L_{max}$ and the lower bound $m_{min}$ is given by the mass
of the first excited state in $N$ channel, with unphysical quark mass properly
taken into account by (\ref{b_mass}).

\begin{figure}[!t]
\vspace{0.75cm}
\includegraphics[width=6.5cm]{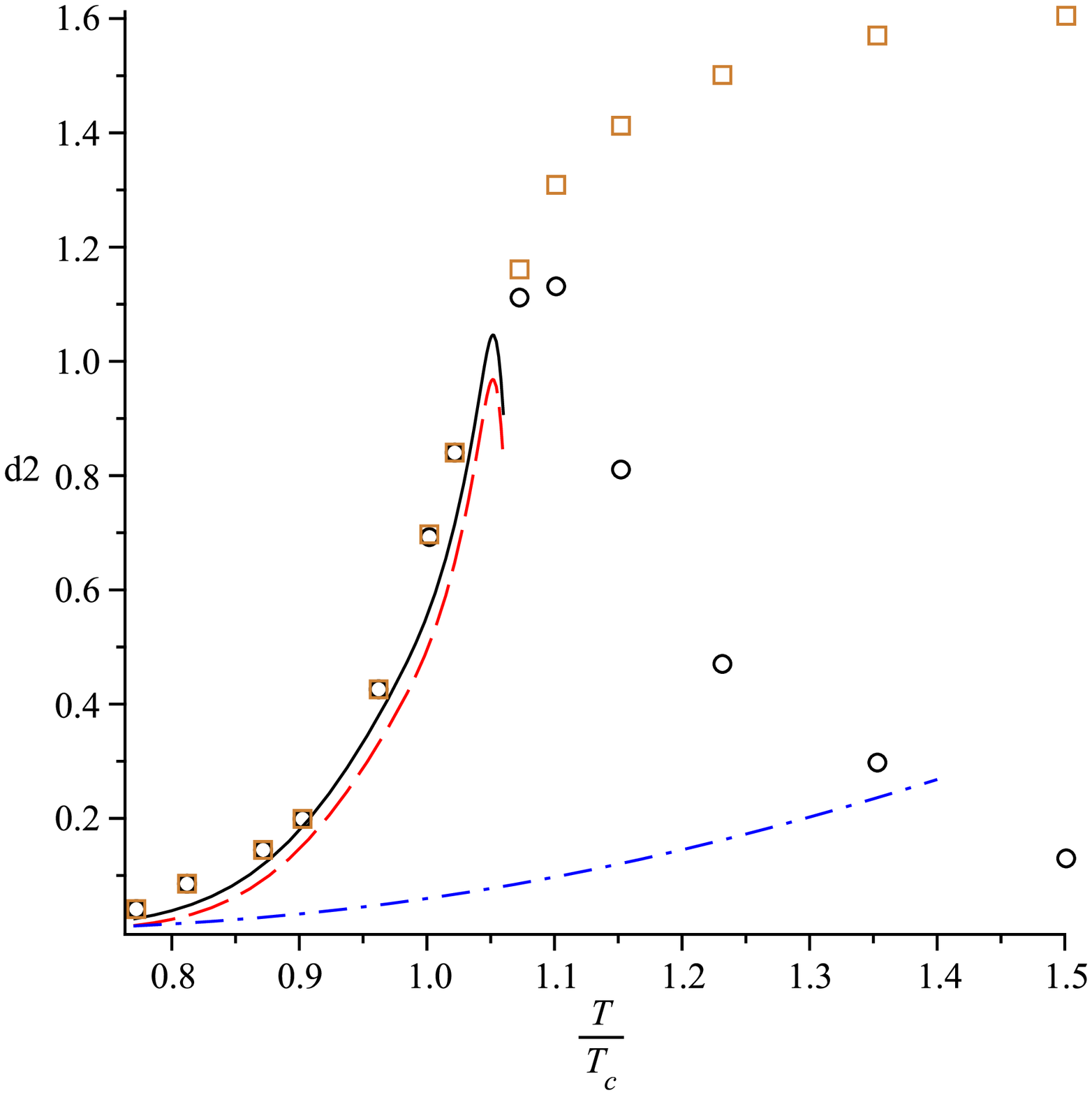}\\
\includegraphics[width=6.5cm]{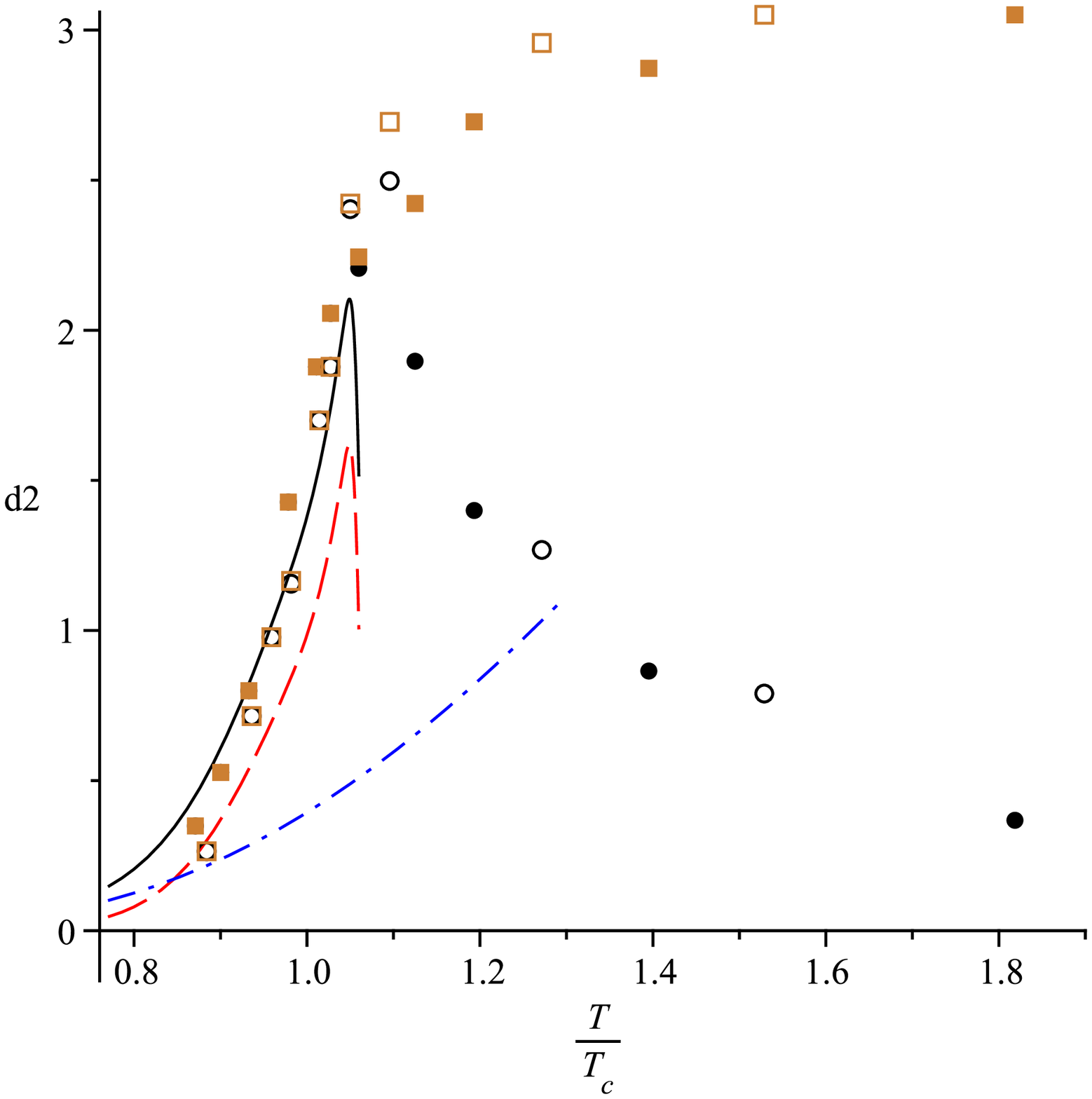}
\caption[h]{\label{fig_dn_str}(color online) Various contributions
to the susceptibility $d_2(T)$ as a function of the temperature. 
The upper figure shows the
lattice data of the Set 1(2 flavors at $m_{\pi}=770 MeV$) from ref.\cite{allton}, with quark contribution subtracted, are shown 
by (black) circles. The unsubtracted data are also included, as (brown) boxes
for comparison.)
The (red) dashed line corresponds to 
hybrid (or ``stringy'')   baryons. The (blue) dash-dotted line corresponds 
to the sum of the lowest 20-plet states. 
The (black) solid line is the overall baryonic contribution.
The parameters used are $x=0.62$, $y=0.08$.
The lower plot is for  Set 2 (2+1 flavors at $m_{\pi}=220 MeV$)
\cite{cheng}.
The lattice data on 
susceptibility\cite{cheng} with quark contribution subtracted are shown
as black circles(solid ones for $N_\tau=6$ and open ones for $N_\tau=4$).
The unsubtracted data are also included as brown boxes
(solid ones for $N_\tau=6$ and open ones for $N_\tau=4$) for comparison.
 The parameters used are $x=0.51$, $y=0.08$. }
\end{figure}

Now is the time to show the results and confront it with lattice findings
 on susceptibilities. We show those, with and without
the free quark contribution, 
%
%
in Fig.\ref{fig_dn_str} for the first susceptibility $d_2$. 

The red dashed line shows the contribution of the ``stringy'' baryons, while the dash-dotted line shows
the contribution of the lowest ``chiral'' states. 
First of all,  the  stringy states, 
albeit higher in mass, predominates over the lowest states near $T_c$, 
and their rise explains the data, both in magnitude and in the growth rate.   Their dominance is more prominent
at larger bare quark mass(Set 1) than at smaller bare quark mass
(Set 2), the latter having smaller masses for the lowest states. Another reason why Set 2 has larger contribution
of the non-stringy baryons is that it has $N_f=2+1$ rather than 2, causing further proliferation of baryons relative to quarks.

As the temperature 
rises above $T_c$, the string model predicts the fall of the baryons caused by quark proliferation and a decrease
of the string length. It happens  when $m_{min}\sim m_{max}$ which occurs
at $T\sim 1.05T_c$. This value is in qualitative agreement with the result
obtained from the $L_{max}\sim a$ condition in the previous section. 
Unfortunately, this fall is
much more rapid than the data, and the question arises what is the physical nature of states between 1.1-1.4 $T_c$ for set 1 and
1.1-1.8 $T_c$ for Set 2.

Those cannot be the ``non-stringy'' baryons, shown by the dash-dotted curves. First of all, their contribution have the wrong shape:
it is increasing rather than decreasing. Furthermore, we terminated those lines at  the
 melting temperature of baryons
  calculated in \cite{Liao:2005hj}, modified for the particular quark masses
used here. 



\section{Chiral dynamics and the baryons} \label{sec_chiral}

\subsection{Why do we have two types of hadrons?}
    In a narrow sense, chiral dynamics is related to the dependence on the dynamical quark masses in the theory $m_q$ 
when they are small or vanishing. In a more broad sense the issue includes an old question of to what extent the baryons are made   
of  their ``pion cloud'', and to what extent of quarks and confining strings, as in the nonrelativistic quark model.
 Hadronic physics has been oscillating between these two conflicting points of view for decades, 
 with drastically different expectations for $T$ and $\mu$ dependencies of hadronic masses following from them.

One extreme point of view, following from the Skyrme model and application of the QCD sum rules, 
suggested that hadronic masses are due to sigma/pion bags and are impossible without the chiral symmetry breaking and nonzero $\langle\bar{\psi}\psi\rangle$. If so, one would expect drastic changes or even vanishing of hadronic masses at the chiral restoration transitions  
 \cite{BrownRho}. 
 
 The opposite  point of view is that the sensitivity of baryon masses to chiral dynamics is restricted to
 explicit chirally odd coefficient of the quark mass, the sigma term. Here we will adhere to this conservative point of view.
%
%
%
%

Before turning to the chiral effect on the baryon mass and density
of states, we want to
point out a significant difference between (i) the lowest baryons (in each channel, such as $N,\Delta$) and (ii) the bulk of hybrid baryons.
  
  The former ones (i) -- as experiment and lattice simulations told us -- are strongly interacting with pions and chiral condensate. 
  They do not have nearby parity partners.
   Significant part of their
  mass is believed to be related to ``the pion cloud'', the extreme model on that is of course the Skyrme model which look at  them  as being pionic solitons. 
  Instanton liquid model explains their masses well, using propagators based on instanton zero modes.
  
  The latter (ii), the bulk of excited baryons, are on the contrary very insensitive to chiral physics. The most vivid spectroscopic evidence for that
  is parity doubling phenomenon, see extensive set of data on that in \cite{Glozman}.  As was repeatedly emphasized in discussions of the matter,
  close pairs of opposite parity also imply weak coupling to pions, which was also evident from the magnitude of decay matrix elements. 
  
  Why are these excited baryons behave like this? Our interpretation of them as ``stringy'' excitations would suggest that perhaps interaction of the
  color string and chiral condensate is small. This was in fact demonstrated on the lattice already 20 years ago \cite{Feilmair:1989hi}: in the vicinity
  of static quark the $\langle\bar\psi\psi\rangle$ is reduced by few percents only.  
  
  Instanton liquid model provides another interpretation of this phenomenon: it is a reflection of the Dirac spectrum in the field of an instanton,
  in which there is one (chiral) zero mode plus a continuum of non-chiral nonzero ones. The lowest hadrons are in this model
  viewed as exceptional states, built from collectivized zero modes, unlike all the excited states.

The lowest excitations are described via some potential models (especially accurate for heavy quarkonia) as radial and orbital excitations
of moving quarks. Excitations in which quarks are basically at the same state as in the lowest hadrons but a $string$ is excited are called  {\em ``hybrid states''}:
there is vast lattice and experimental literature about them. The
string-Hagedorn argument mentioned above implies, that such states are dominant at large excitations.



\subsection{Chiral restoration transition and the lowest baryons }

\begin{figure}[!t]
\vspace{0.75cm}
\includegraphics[height=5cm,width=5cm]{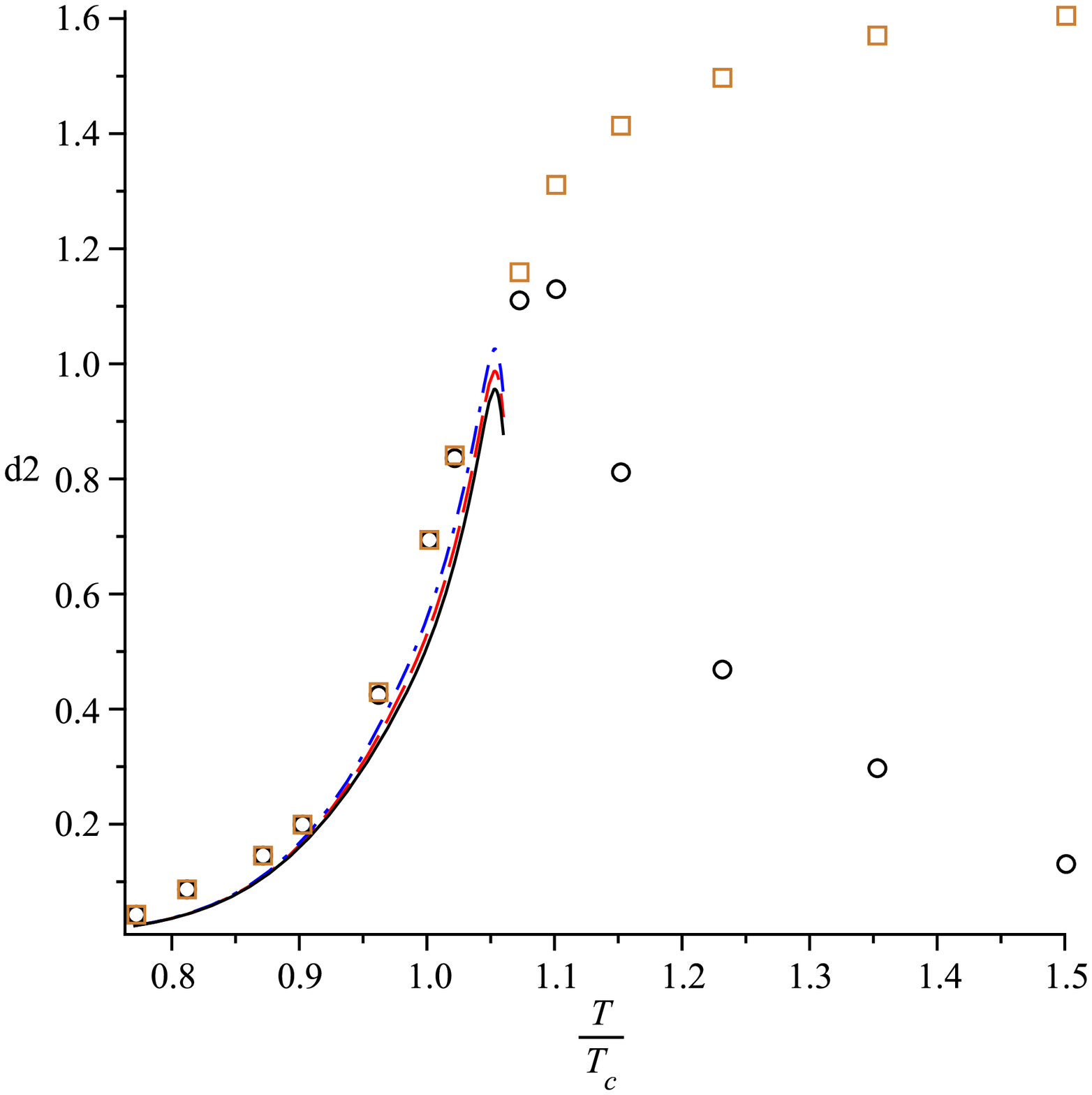}
\includegraphics[height=5cm,width=5cm]{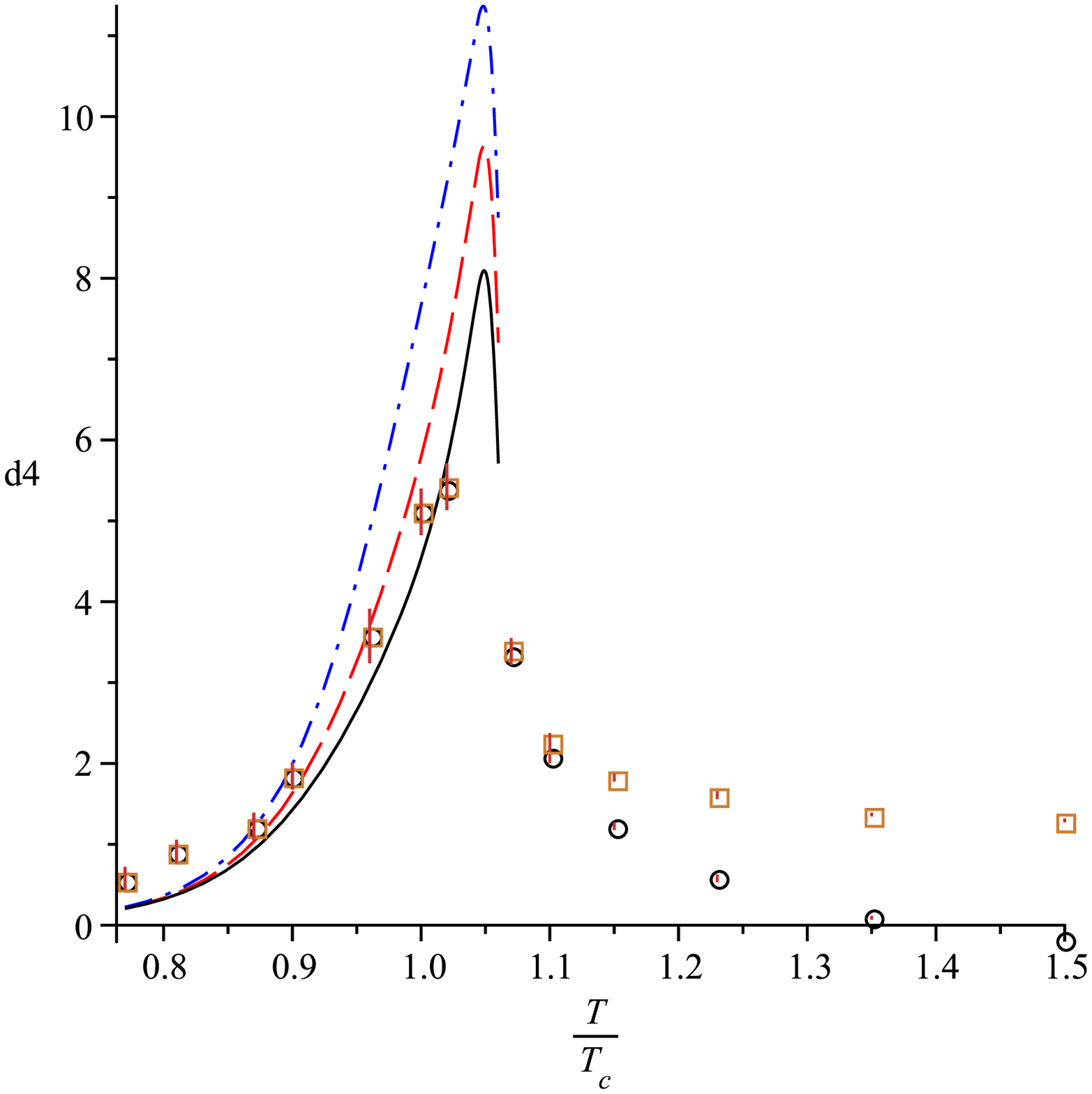}
\includegraphics[height=5cm,width=5cm]{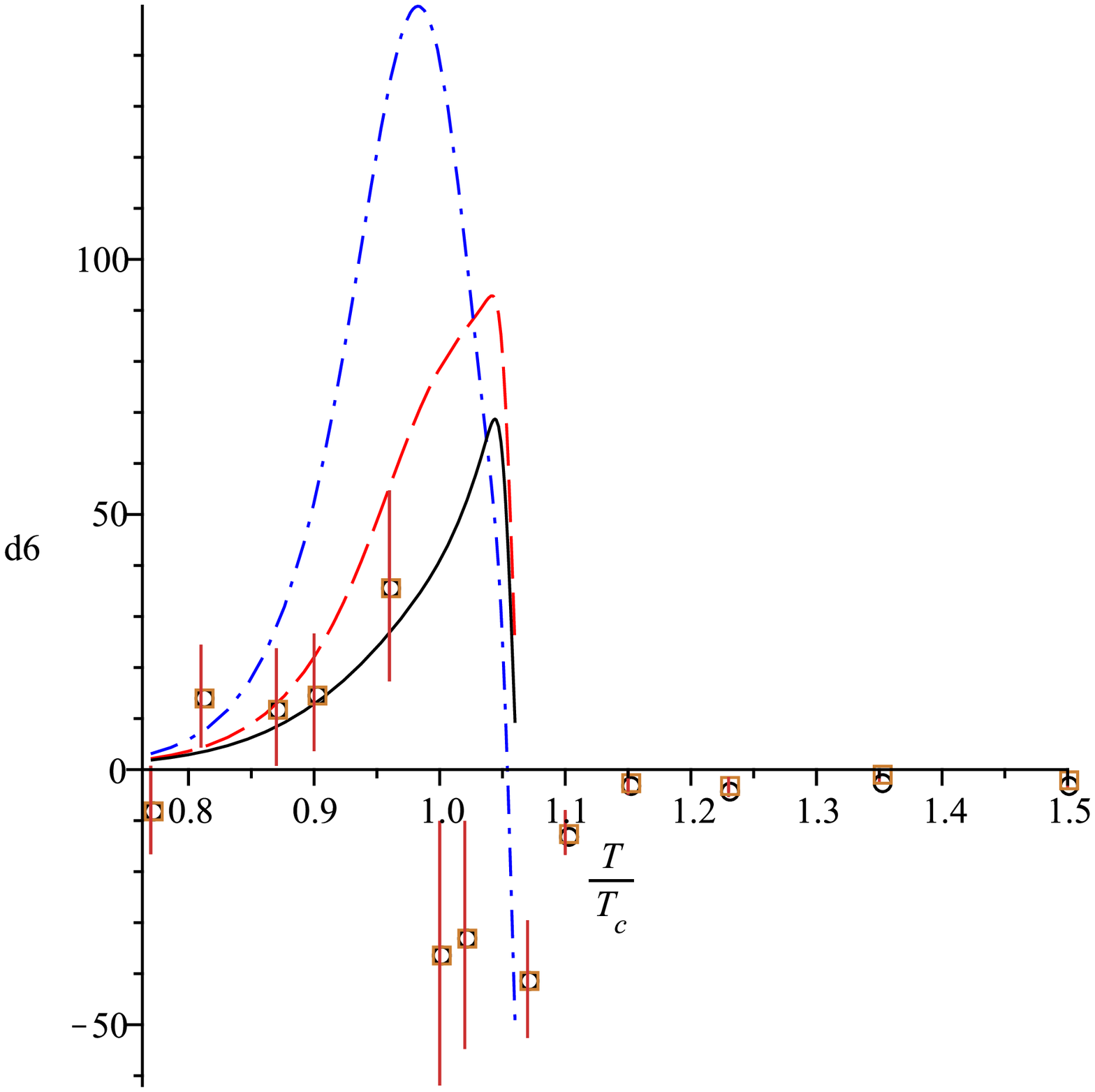}
\caption[h]{\label{fig_dn}(color online)Three figures display the temperature dependence of the second, fourth and sixth susceptibilities.
The points are the Set 1 lattice data 
 with
 the quark contribution subtracted. 
The lines correspond to the model discussed in the text, with $\mu$ dependent 
stringy and chiral baryons. The parameters used are $x=0.6$, $y=0.08$,
and $\mu_c=\infty$(black solid line), $\mu_c=2T_c$(red dashed line) and
 $\mu_c=1.33T_c$(blue dash-dotted line).}
\end{figure}

\begin{figure}[!ht]
\vspace{0.75cm}
\includegraphics[width=5cm]{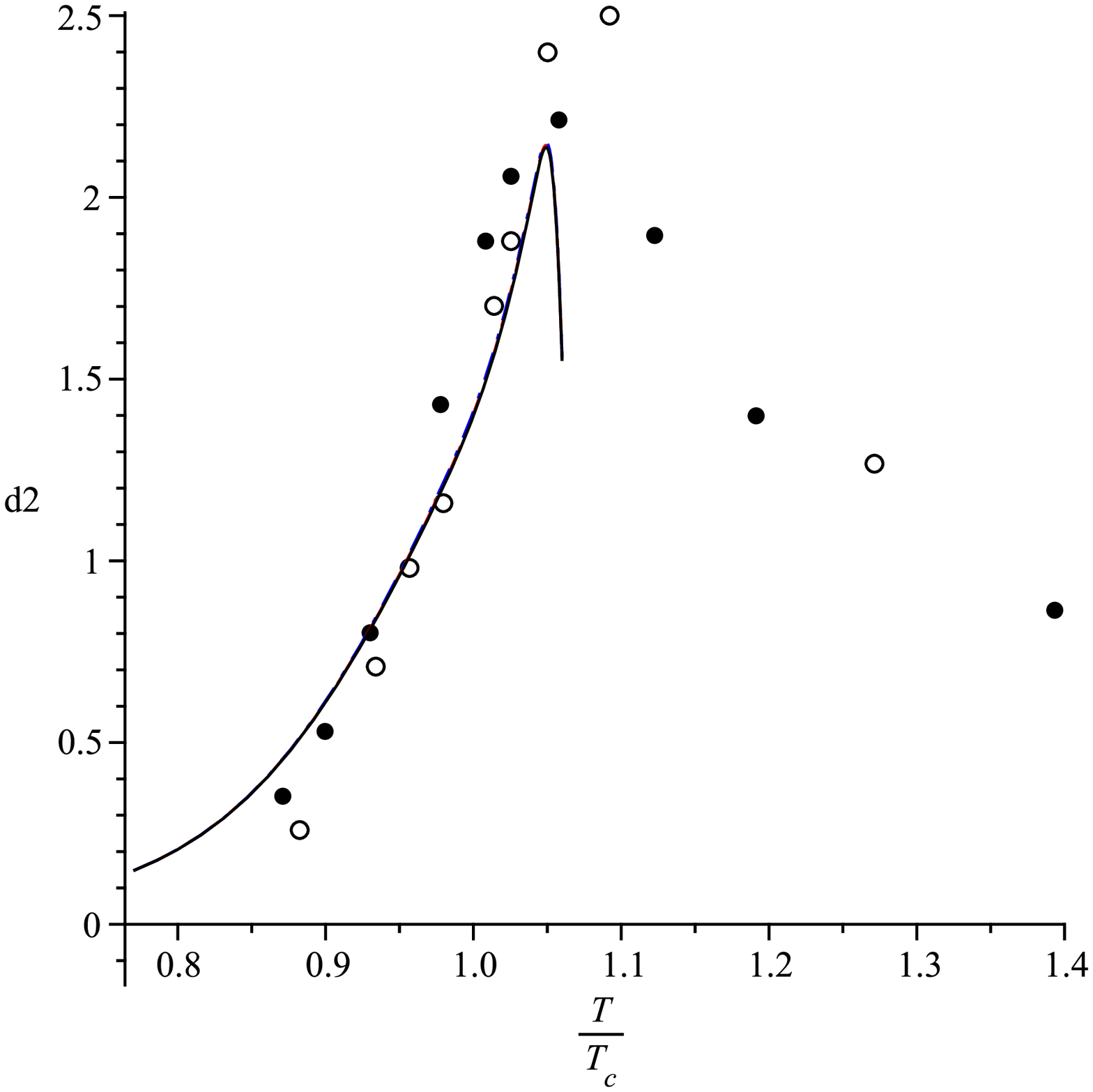}
\includegraphics[width=5cm]{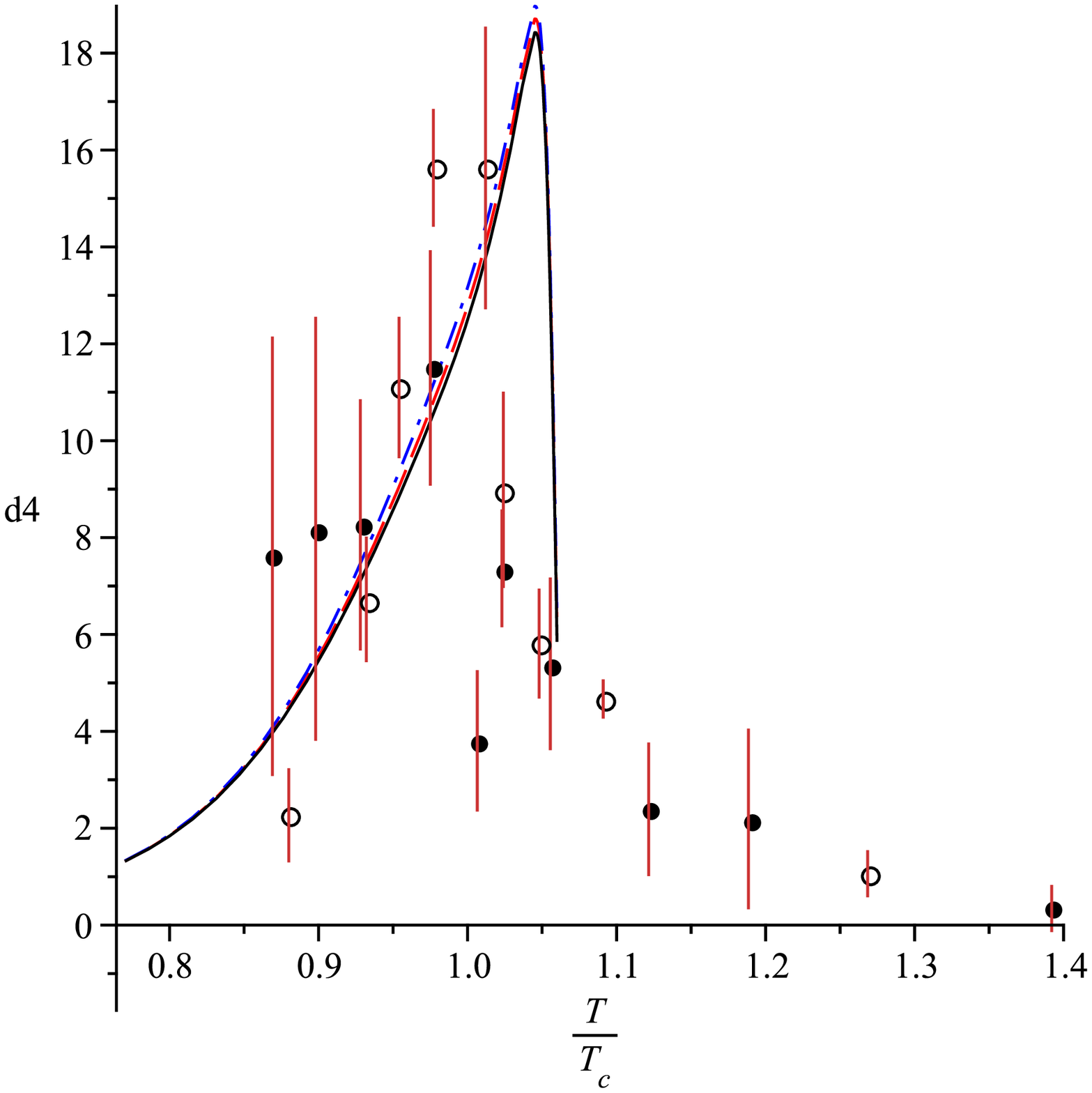}
\caption[h]{\label{fig_dn_nf3}(color online)
The temperature dependence of the second and  fourth  susceptibilities, for Set 2
lattice data, with the quark contribution subtracted.
 The (black) solid circles correspond to $N_\tau=6$ 
and open circles to $N_\tau=4$.
The curves are for the model described in the text, with
 parameters values equal to $x=0.51$, $y=0.08$,
and $\mu_c=\infty$(black solid line), $\mu_c=2T_c$(red dashed line) and
 $\mu_c=1.33T_c$(blue dash-dotted line).}
\end{figure}

 In the model proposed we  assumed that the strongest sensitivity to small quark mass is entirely due to the ``lowest'' states
  with each quantum numbers. 
 The mass of some baryonic
state can be expressed by
\be
m_B=m_q\langle B|\bar\psi\psi|B\rangle+\text{``gluonic part''}
\ee
where the coefficient of the quark mass is the scalar quark density: it is known as the sigma term.

In the calculation above we had fixed its magnitude from lattice works, in the vacuum: now is the time to discuss
its dependence on the temperature.
Following the idea that any chirally-odd quantity should disappear at chiral restoration transition, and furthermore that
the sigma terms for any hadron is more or less proportional to the VEV or quark condensate,
 we propose to incorporate the chiral effect and $\mu$
dependence through the following formula:

\be
m(T,\mu)=km_{\pi}^2\frac{\langle B|\bar\psi\psi|B\rangle(T,\mu)}{\langle B|\bar\psi\psi|B\rangle|_{T=\mu=0}}+ N_s m_s+\sigma_V(T)L(T)
\ee

In the absence of knowledge of the expectation value of $\bar\psi\psi$ with
respect to the baryonic state, we simply assume normalized chiral condensate
associated with some baryonic state is given by its counterpart in vacuum state:
\be
\frac{\langle B|\bar\psi\psi|B\rangle(T,\mu)}{\langle B|\bar\psi\psi|B\rangle|_{T=\mu=0}}=\frac{\langle\bar\psi\psi\rangle(T,\mu)}{\langle\bar\psi\psi\rangle|_{T=\mu=0}}
\ee

The temperature dependence has been measured on the lattice\cite{petreczky}.
We fit the normalized chiral condensate by the following:
\be\label{cc}
\frac{\langle\bar\psi\psi\rangle(T,\mu=0)}{\langle\bar\psi\psi\rangle|_{T=\mu=0}}=0.55-
0.42\tanh(9.47\frac{T}{T_c}-9.55)
\ee

The $\mu$ dependent term enters quadratically by the
 $\mu\leftrightarrow -\mu$ symmetry. We will use the collective coordinate
$\sqrt{(T/T_c)^2+(\mu/\mu_c)^2}$ instead of $T/T_c$. Similar parametrization is
also used in \cite{Liao}. We leave $\mu_c$ as a free parameter.
With chiral effect incorporating both $T$ and $\mu$ dependence, 
we are ready to calculate the susceptibility
using (\ref{string_dn}). Since $\mu$ appears in many places in (\ref{string_p}),
taking derivatives becomes cumbersome. We simply use numerical differentiation
method. Further simplification can be made by noting the evenness of 
(\ref{string_p}) as a function of $\mu/T$. $2n$-th derivative with respect to
 $\mu/T$ amounts to $n$-th derivative with respect to $(\mu/T)^2$.

As already discussed in \cite{Liao,Karsch_qm}, in the model of
baryon gas the deviations of $d_4/d_2$ from $9$ follows 
from the fact that their masses
are effectively dependent on $\mu$. In our string model, this appears as
a $\mu$ dependent density of states $\rho(T,\mu)$\footnote{We of course
also have $\mu$ dependence in the mass upper cutoff of the hybrid states, but
it only causes small deviation}.

We show the effect of $\mu$ dependence of baryon mass in 
Fig.\ref{fig_dn}(Set 1) and Fig.\ref{fig_dn_nf3}(Set 2). We see
the inclusion of this effect hardly
change the $d_2$ plot, but shifts the peak upwards
 in $d_4$ and $d_6$ plots. Smaller values of $\mu_c$ result in higher peaks.
However the peaks of $d_4$ and $d_6$ are over predicted either with($\mu_c=1.33,2T_c$)
 or without($\mu_c=\infty$) $\mu$ dependence of the baryon mass. 
The shift of the peak is more prominent near $T_c$, where the 
chiral condensate drops rapidly, thus the masses of the lowest states 
change rapidly as well. It is also worth noting that the shift is less
prominent in case of 2+1 flavors model. This is because in that case 
a smaller fraction 
of the mass has changed due to a smaller sigma term.

It is interesting to note that the peaks of susceptibility data(with quark
contribution subtracted) move to lower temperature as we go to higher 
cumulants, e.g. $d_2$ in Fig.\ref{fig_dn} peaks at $T\sim 1.1T_c$ while
$d_6$ peaks at $T\sim 1.0T_c$. This might be an artifact of our simplified
treatment of quark contribution (\ref{p_quark}). It is worth mentioning that
 a recent study with holographic model\cite{holo} shows nontrivial
temperature dependence of susceptibilities of different cumulant, which deserves
further understanding.

\begin{figure}[!ht]
\vspace{0.75cm}
\includegraphics[width=7cm]{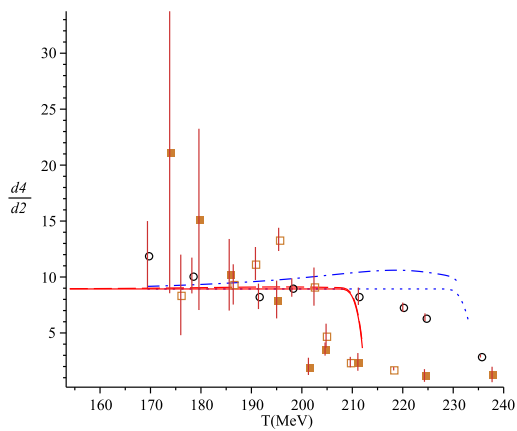}
\vspace{0.75cm}
\caption[h]{\label{fig_kur_mq}(color online)The temperature dependence of the kurtosis $\frac{d_4}{d_2}$
All discussed
lattice data are shown here:
Set 1 (2 flavor at $m_{\pi}=770MeV$) by (blue) circles, Set 2  (2+1 flavor at $m_{\pi}=220MeV$) by
(red) solid boxes and open boxes, for $N_\tau=6$ and $N_\tau=4$
respectively. 
Four curves  in the plot are for  2 flavor at $m_{\pi}=770MeV$ 
without (blue dotted) and with (blue dash-dotted) chiral effect, 
2+1 flavor at $m_{\pi}=220MeV$ without (red solid) and with (red dashed)
chiral effect. The same parameters are used as in 
Fig.\ref{fig_dn_str},\ref{fig_dn},\ref{fig_dn_nf3}. In particular, we 
have chosen $\mu_c=2T_c$ for cases with chiral effect) 
}
\end{figure}

It is also instructive to plot the kurtosis from our stringy baryon model.
We show the plot of kurtosis both with and without the chiral effect
 in Fig.\ref{fig_kur_mq}. A sharp dropping of the kurtosis
is observed near the critical temperature. However the limitation of the
model prevents us to go beyond $1.05T_c$, where the kurtosis is supposed to
drop to 1. The importance of $\mu$ dependence of the baryon mass is clearly
illustrated in Fig.\ref{fig_kur_mq}. Without the $\mu$ dependent density of
states, the drop is a simple vertical fall, while the inclusion
of chiral effect produces a bump where the chiral condensate drops rapidly.
The bump is more prominent for 2 flavors Set 1, in which case the lowest
states have relative large fraction of mass that is chirally sensitive. 

The sign and magnitude of the chiral restoration effect in our model  is in
sharp contrast with the PQM model 
\cite{stokic}, predicting sharp and very large  deviation of the
kurtosis from $9$, up to about 15. That happened because 
in Brown-Rho-type dropping of  hadronic masses prior to chiral restoration transition.
In Fig.\ref{fig_kur_mq} there is certain mixed signal here:
Set 2 data at $N_\tau=4$ and $N_\tau=6$ show somewhat different trend. Unfortunately
current data are still insufficient to make clear distinction between our conservative approach
(only the sigma terms melt at chiral restoration) and the Brown-Rho behavior. We hope this interesting
issues will be clarified in the future.

\section{Discussion}\label{sec_discussion}
\subsection{Quarks bound to monopoles?}

As we have already emphasized above, lattice data on the deconfinement transition give rather different width of the transition region for baryons
(discussed in the preceding section) and the quarks. While the baryons show rapid rise and then rapid fall already at about $1.05T_c$, quark suppression is dominated by the Polyakov loop, which only goes to 1 at $2T_c$. Such mismatch of the widths, be nearly and order of magnitude, leave some ``hole'' in the middle: indeed,
 one can see from Fig.\ref{fig_dn} that our model has certain deficit at $T= (1.1-1.4) T_c$, as compared to the lattice data. So,
 it can be a deficiency of the model, or some missing component.
 
 Thinking about possible missing states leads us to propose one  idea, which is by itself well known in theoretical literature, yet (for our knowledge) never been used in QCD
 thermodynamics context so far. Static monopole solution has 3-dimensional  fermionic zero mode. It is important to note, that its existence follows from topological theorems
 and thus does not depend on details of the monopole profile. In particular, it exists whether monopoles can or cannot be described semiclassically. 
 
  In vacuum $T=0$ this simply means that 
a quark can be bound to the monopole. For a massless quark in vacuum  
 there are two degenerate
 states, with fermionic numbers 1 and 0\footnote{ in the context of $SU(2)$
supersymmetric model those are sometimes redefined and called states with
 fermionic number 1/2 and -1/2: but we would not do so. Finite temperature breaks the supersymmetry anyway.}. For $N_f$ quark flavors
 each can be either present or absent, leading to $2^{N_f}$ states.
 
 Euclidean finite $T$ formalism breaks the degeneracy of this situation:  the  bosonic objects should have periodic wave function, and thus time-independence is unaffected,
 while fermionic objects should have antiperiodic wave functions in Euclidean time. Thus one should inevitably introduce $exp( i E\tau)$ with
 the lowest odd Matsubara frequencies $E=\pm \pi T$ into the fermionic wave function, making the composite fermionic magnetic object heavier
 \be M_{fm}\approx M_{mono}+\pi T\ee 
 The mass of the monopole is believed to be minimal at $T_c$, while its value is  not yet measured. It has been argued from BEC condition \cite{Cristoforetti} 
     that it may be as small as $M_{mono}\sim 200 \, MeV$: if so, the minimal mass of the composite object is about $ M_{fm}(T=T_c)\sim .8\, GeV$,
     which is accidentally close to lattice data on quark quasiparticles. Note however, that such composite has only $one$ spin state.
     
     At temperatures $T\sim T_c$, the monopoles remain strongly correlated
and behaves as liquid\cite{liquid}, preventing us using ideal gas formula. 
The scaled monopole density is measured in $SU(2)$
gauge theory on the lattice\cite{mono_density} and is
parametrized by:

\be
&&\frac{n}{T^3}=\frac{A}{ln^2(\frac{T}{\Lambda})}\\
&&\text{with}\;A=0.557\;\frac{T_c}{\Lambda}=2.69
\ee

\begin{figure}[!ht]
\vspace{0.75cm}
\includegraphics[width=8cm]{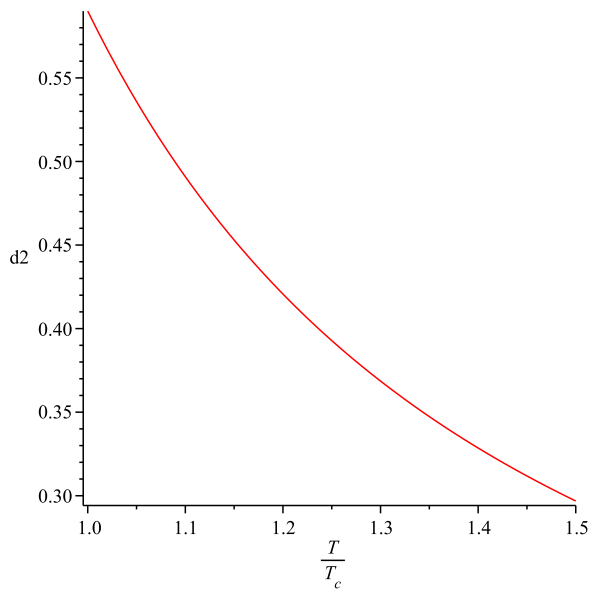}
\caption[h]{\label{fig_mono}The contribution from
monopole(antimonopole)-zero mode bound state. We have used a degeneracy factor
$2N_cN_f$ since the zero mode only has one spin}
\end{figure}

The density of monopole(anti-monopole)-quark bound state is estimated as
$n e^{-\pi}$. The additional Boltzmann factor arises from the Matsubara frequency.
We plot the bound state contribution in Fig.\ref{fig_mono}.
    Taking into account a very schematic model and preliminary data on the monopole density, we conclude that magnetic composite fermions
    do have a potential to be the ``missing ingredient'' mentioned at the beginning of this section. Of course, to know it for sure much more work is needed.
    In particularly, existence of such composite fermions with a magnetic charge can be investigated on the lattice, by
observing the  correlations between the monopole paths and the fermionic operators (especially the fermionic spin).
    
\subsection{ Outlook}\label{sec_outlook}
We hope that multiple details of the last sections have not detracted the reader from the main ideas
of the model, which are (i) exponential rise of the stringy partition function near $T_c$, as well as (ii)
the upper cutoff of the strings length defined by the quark density. Those two phenomena are the basis for
``the rise and fall of the baryons'' put into the title.

Our conservative treatment of the chiral dynamics seems to be working.
The dependence of the baryon masses on $m_q$, light quark bare mass, and related effects due to chiral symmetry restoration
and vanishing of the ``pion cloud'' are small but interesting corrections to the main picture. We have tried -- using lattice
data whenever those are available -- to parametrize those effects and include them into the model, in the last few sections.
The reader  may at this point ask what is actually achieved by doing this. 

Any model is as good as ideas put into it, and in order to test it one can see if it has any predictive power. By this we mean
its application to wide range of the QCD-like theories. So far we considered two directions: (i) variable light quark mass, and (ii)
increasing  $\mu/T$ or -which is basically the same -- including the susceptibilities with more derivatives.

 We suggest to address the same issues using other theories, especially 
 with dynamical $adjoint$ fermions. In this case the critical points of the deconfinement transition
and that of chiral restoration are very different    
\be T_{deconfinement} \ll T_{chiral}\ee
so that there is a new intermediate phase, that of ``deconfined constituent quarks''. 
For three colors the deconfinement is in this case the first order transition, while the chiral one is of the second order, see
e.g.  \cite{Cossu:2008wh} and references for earlier papers therein.
Another reason one may be interested in this direction is possible relation to supersymmetric theories.

For simplicity and continuity of the observables, one may
introduce non-dynamical fundamental  quarks which would have their own  chemical potential and form baryons with the usual
quantum numbers, basically repeating the same analysis as was done for Set.1 and 2 we used , but 
with a different ensemble of gauge fields, now with some dynamical adjoint fermions.
 In this case one  expect to find the ``rise and fall'' of the stringy baryons be seen at the deconfinement transition, while the effects related with chiral condensate
(and sigma term) melting 
occurring at completely different chiral restoration temperature. Large intermediate phase would be a great place to finally find out whether confinement
is or is not required for the existence of the ``chiral'' baryons.
Another difference between adjoint and fundamental fermions is a different number of fermionic zero modes of monopoles/dyons: there are $two$
bound states of adjoint fermions rather than one. So, if such states play a role above the deconfinement, they should be significantly more
important in the adjoint theories, especially those with more than one fermion flavors.
 
%

\section*{ Appendix}

\begin{appendix}
In this appendix we collected some information we used from different lattice works, to parametrize several
quantities of interest.

\begin{figure}[!ht]
\vspace{0.75cm}
\includegraphics[width=6cm]{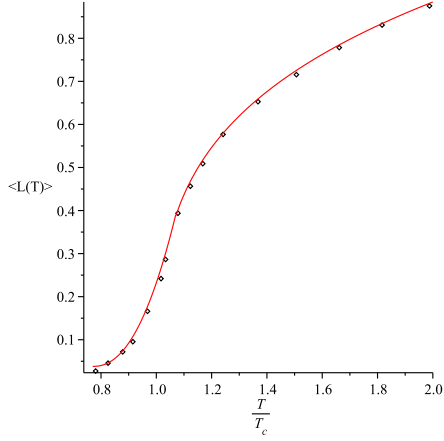}
\caption[h]{\label{fig_polyakov}(Color online) The points are lattice data on the  temperature dependence of the Polyakov loop 
from Ref.\cite{KZ},  fitted with the expression (\ref{fit_poly})}
\end{figure}

{\bf The Polyakov loop} in two flavors 
QCD is shown in Fig.\ref{fig_polyakov}. It reaches 1 at 
around $2.5T_c$, where the quark density reduces to the free quark situation. 
The Polyakov loop is parametrized by:
\be\label{fit_poly}
\langle L(T)\rangle=\left\{\begin{array}{l@{\quad\quad}l}
0.038+5.20(T-0.77)^{2.24}& T<1.07T_c\\
-0.164+1.048(T-1.0)^{0.242}& T>1.07T_c
\end{array}
\right.
\ee

\begin{figure}[!ht]
\vspace{0.75cm}
\includegraphics[width=6cm]{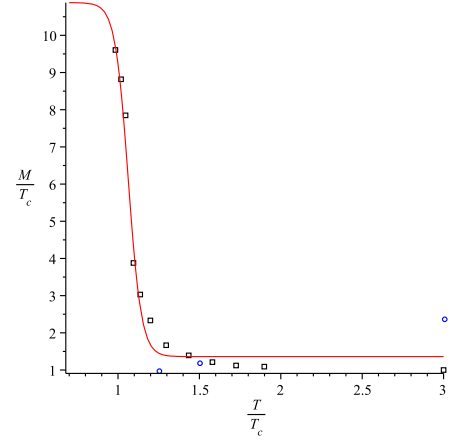}
\caption[h]{\label{fig_mq} (Color online) The points are lattice data on the  temperature dependence of the effective quark mass,  deduced from
the potentials studied in  Ref.\cite{KZ2}.
The (red) line is a fit with expression (\ref{quasi_mass}). Direct measurements of the quark effective mass from the dispersion relations of the quark quasiparticles
for two temperatures, $T=1.25T_c,1.5T_c,3T_c$, from Ref.\cite{quark_quasi} 
are  shown by (blue) circles  for comparison.}
\end{figure}

 {\bf The thermal quark mass} as a function of the temperature is shown  in Fig.\ref{fig_mq}. The thermal quark mass $m_q$ around the critical temperature is important as it affects the cutoff in 
the baryon density of states via $n_q$. Recent lattice 
measurement\cite{quark_quasi} indicated $m_q/T\sim 0.8$ in temperature range
$T\sim 1.25-3T_c$. Thermal mass below $1.25T_c$ is still not known. In the
absence of additional data, we estimate the quark thermal mass as half of the
internal energy for quark-antiquark at infinity separation $U_\infty$. The data
of $U_\infty$ are taken from Table I of \cite{KZ2}\footnote{We use data only at
temperature above the critical value. Below the critical temperature, half
of $U_\infty$ gives the meson mass, and thermal quark mass is not important 
due to Polyakov loop suppression}. The thermal mass of the quarks are fitted
with:

\be\label{quasi_mass}
\frac{m_q}{T_c}=6.12-4.76\tanh(12.38\frac{T}{T_c}-13.16)
\ee

\begin{figure}[!ht]
\vspace{0.75cm}
\includegraphics[width=6cm]{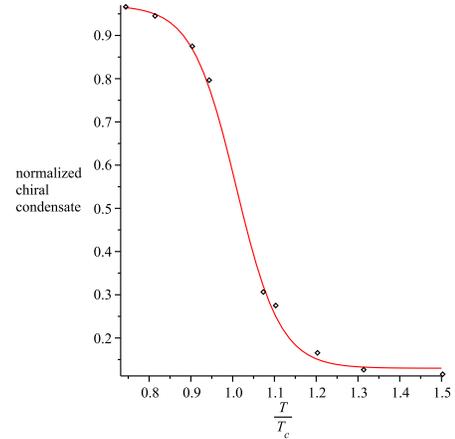}
\caption[h]{\label{fig_cc} (Color online)The temperature dependence of the chiral condensate, normalized to its vacuum value. 
The (black) dots are taken from lattice work Ref.\cite{petreczky} and fitted with 
the red curve  according to expression (\ref{cc}).}
\end{figure}

{\bf The chiral condensate} from the lattice measurement 
(\ref{cc}).
is shown 
In Fig.\ref{fig_cc}.  It is defined as, $\frac{\langle\bar\psi\psi\rangle_{l,T}-\frac{m_l}{m_s}\langle\bar\psi\psi\rangle_{s,T}}{\langle\bar\psi\psi\rangle_{l,0}-\frac{m_l}{m_s}\langle\bar\psi\psi\rangle_{s,0}}$ and measure at finite 
mass ratio of light and strange quark $\frac{m_l}{m_s}$. We will however use
this value as our normalized chiral condensate, which is valid at $\frac{m_l}{m_s}\rightarrow 0$

We fix the $\mu$ dependence by the following ansatz:

\be
&&\frac{\langle\bar\psi\psi\rangle(T,\mu)}{\langle\bar\psi\psi\rangle|_{T=\mu=0}}= 0.55-\nonumber\\
&&0.42\tanh(9.47\sqrt{(T/T_c)^2+(\mu/\mu_c)^2}-9.55)
\ee

\end{appendix}
\section* {Acknowledgments}

We thank Jinfeng Liao, Chuan Miao and Claudia Ratti for helpful discussions. 
Our work was partially supported by the US-DOE grants DE-FG02-88ER40388 and
DE-FG03-97ER4014.

\end{document}